\documentstyle{mn}
\def\be{\begin{equation}}
\def\ee{\end{equation}}

\clearpage
\newpage

\title{The dynamics of S0 galaxies and their Tully-Fisher relation}

\author[A.~Mathieu, M.R.~Merrifield and K.~Kuijken]{A.~Mathieu,$^1$
M.R.~Merrifield$^1$ and K.~Kuijken$^2$\\ 
$^1$School of Physics and Astronomy, University of Nottingham, 
    Nottingham NG7 2RD\\
$^2$Kapteyn Institute, Postbus 800, Groningen 9700 AV, the Netherlands}
\date{Received date; accepted date}

\begin{document}
\label{firstpage}
\maketitle

\begin{abstract}
This paper investigates the detailed dynamical properties of a
relatively homogeneous sample of disc-dominated S0 galaxies, with a
view to understanding their formation, evolution and structure.  By
using high signal-to-noise ratio long-slit spectra of edge-on systems,
we have been able to reconstruct the complete line-of-sight velocity
distributions of stars along the galaxies' major axes.  From these
data, we have derived both model distribution functions (the phase
density of their stars) and the approximate form of their gravitational
potentials.

The derived distribution functions are all consistent with these
galaxies being simple disc systems, with no evidence for a complex
formation history.  Essentially no correlation is found between the
characteristic mass scale-lengths and the photometric scale-lengths in
these galaxies, suggesting that they are dark-matter dominated even in
their inner parts.  Similarly, no correlation is found between the
mass scale-lengths and asymptotic rotation speed, implying a wide
range of dark matter halo properties.

By comparing their asymptotic rotation speeds with their absolute
magnitudes, we find that these S0 galaxies are systematically offset
from the Tully-Fisher relation for later-type galaxies.  The offset in
luminosity is what one would expect if star formation had been
suddenly switched off a few Gyrs ago, consistent with a simple picture
in which these S0s were created from ordinary later-type spirals which
were stripped of their star-forming ISM when they encountered a dense
cluster environment.

\end{abstract}


\section{Introduction}
Since S0 galaxies seem to have properties intermediate between
elliptical and spiral galaxies, Hubble (1936) placed these gas-poor
disc systems between the spirals and ellipticals, at the bifurcation
of his tuning fork galaxy classification scheme.  However,
three-quarters of a century later, it remains an open question as to
whether this arrangement in any way reflects the physical origins of
S0 galaxies.

One approach to addressing question of this kind has been to search
directly for signs of evolution in galaxy morphologies by observing
samples of these systems over a wide range of redshifts.  In a classic
study of this kind, Butcher \& Oemler (1978) found that the fraction
of blue galaxies in clusters was much higher in the past than it is
now.  They surmised that blue spirals were being converted to
earlier-type systems.  This discovery fitted in extremely well with
Gunn \& Gott's (1972) suggestion that S0 galaxies could form from
spiral galaxies in the dense environment of a cluster via tidal and
ram pressure stripping.  More recently, high-resolution observations
(e.g. Couch et al. 1994; Lavery, Pierce \& McClure 1992) have
confirmed that the blue galaxies in high redshift clusters are,
indeed, relatively normal spiral galaxies, which could well be the
progenitors of S0 systems.

The only problem with such analyses is that one is ne\-ces\-sa\-rily
observing different galaxies in the nearby and distant samples, so any
inference about the evolution from one type to another is of a
circumstantial nature.  A convincing case therefore requires that one
look in some detail at the ``finished articles,'' in order to find any
archaeological evidence for the proposed evolution.  Since galaxies
are intrinsically dynamical entities, one should be able to extract
important clues from their stellar kinematics, as derived from
absorption-line spectra, as well as their photometric pro\-per\-ties.
With high quality spectral data, one can measure the full
line-of-sight velocity distribution (e.g. Koprolin \& Zeilinger 2000,
Fisher 1997), providing information
not only on the motions of the stars, but also the gravitational
potential responsible for the motions.

For S0 galaxies, one important piece of stellar-dynamical evidence
would be provided by an analysis of the Tully-Fisher relation.  In
later-type disc galaxies, there is a strong correlation between
circular rotation speed and optical luminosity (Tully \& Fisher 1977),
and this relationship becomes even tighter when one looks in the near
infrared (Pierce \& Tully 1992).  If S0 galaxies formed in a
relatively benign way from spiral galaxies, one would expect that
their optical luminosities and circular rotation speeds would be
little affected by the process, so a Tully-Fisher relation should
still be apparent; the only significant difference would be that the
stripping of their gas would switch off the star formation process, so
the optical luminosities of the S0s should fade over time, shifting
the zero-point of the relation.

To-date, the most thorough search for a Tully-Fisher relation in S0
galaxies was that made by Neistein {\it et al.}\ (1999).  This
analysis is complicated by the fact that there is no simple measure of
the circular speed in an S0 galaxy: there is no gas at large radii
moving on circular orbits, and the stars follow significantly
elliptical orbits, so their mean streaming velocity at any radius is
lower than the local circular velocity.  Neistein {\it et al.}
overcame the problem by appealing to the equations of galactic
dynamics: by making a few simplifying assumptions, they were able to
use the asymmetric drift equation (Binney \& Tremaine 1987) to combine
the mean streaming velocity and velocity dispersion of the stars in
order to estimate the circular velocity at each radius.  

Although this analysis revealed some evidence for a trend between
I-band luminosity and circular speed, Neisten {\it et al.} found a
huge scatter in the relation, and that there was very little offset in
the mean from the Tully-Fisher relation of later-type galaxies.  They
therefore concluded that these systems could not all have formed from
the simple stripping of spiral galaxies.  Instead, they suggested that
the S0 classification actually represents a rather heterogeneous class
of galaxies, which formed through a rather wide variety of processes.
They further suggested that the absence of an offset in the
Tully-Fisher relation could be understood if the S0 galaxies have more
massive discs, so any fading in their luminosities is offset by the
larger number of stars.

In order to explore these conclusions a little further, this paper
presents a detailed dynamical analysis of six S0 galaxies.  These
galaxies have been selected to contain relatively small central
bulges.  If any S0s formed from simple stripping processes, one would
expect their bulges to be little affected.  Thus, by selecting S0s
with the small bulges characteristic of later-type spirals, one might
hope to pick out a relatively homogeneous subsample of systems that
formed via this route.  Unfortunately, it is only when S0 galaxies are
very close to edge on that one can reliably determine that their
bulges are small.  As Neistein {\it et al.}\ (1999) demonstrated, the
line-of-sight integration of starlight through such an edge-on disc
means that there is quite a large correction to convert the observable
mean line-of-sight velocity into the circular streaming velocity of
the stars.  As a further complication, the observed velocity
distribution for any line of sight through an edge-on disc will be
highly non-Gaussian, due to the contribution from stars at large radii
with small line-of-sight velocities.  A dynamical analysis based on
moments derived from a Gaussian fit to the line-of-sight velocity
distribution is therefore prone to systematic error.  To obviate these
difficulties, the current analysis uses data with a high enough
signal-to-noise ratio for the complete line-of-sight velocity
distribution to be derived, and these data are then fitted to a
complete dynamical model in order to derive both the stellar
distribution function and the gravitational potential needed for the
Tully-Fisher relation.  

The remainder of the paper is laid out as follows.
Section~\ref{datasec} describes the sample, the data analysis, the
model fitting, and the resulting distribution functions and rotation
curves.  Section~\ref{discussionsec} discusses the correlations in
derived quantities, with particular reference to the Tully-Fisher
relation.  Section ~\ref{conclusionsec} presents the conclusions that
can be drawn from this analysis.

\section{Data analysis}\label{datasec}

\subsection{The sample}
\begin{figure}
\vspace*{6.5cm}
\special{hscale=20 vscale=20 hoffset=-20 voffset=50
hsize=500 vsize=800 angle=0 psfile="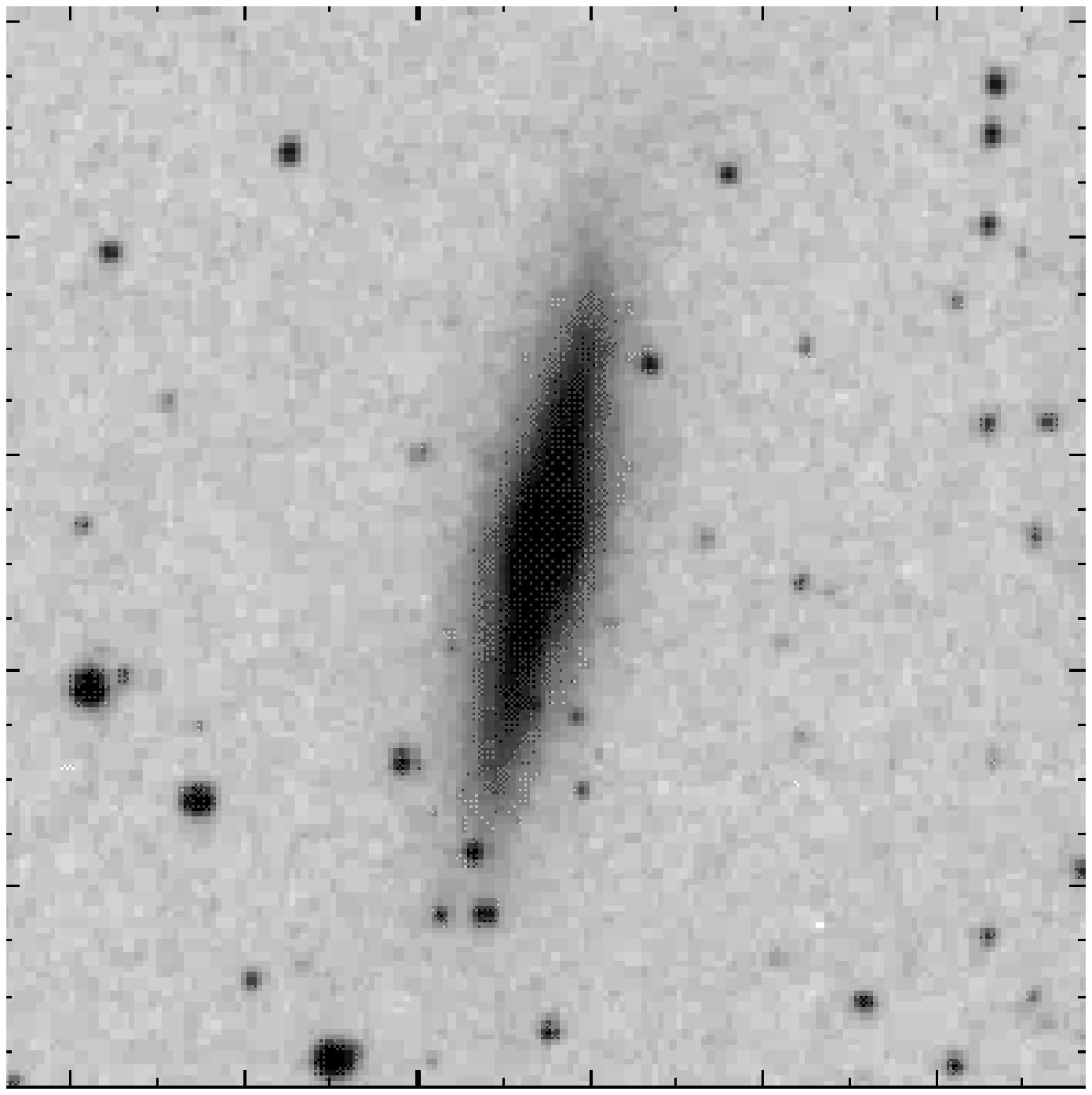"}
\special{hscale=20 vscale=20 hoffset=60 voffset=50
hsize=500 vsize=800 angle=0 psfile="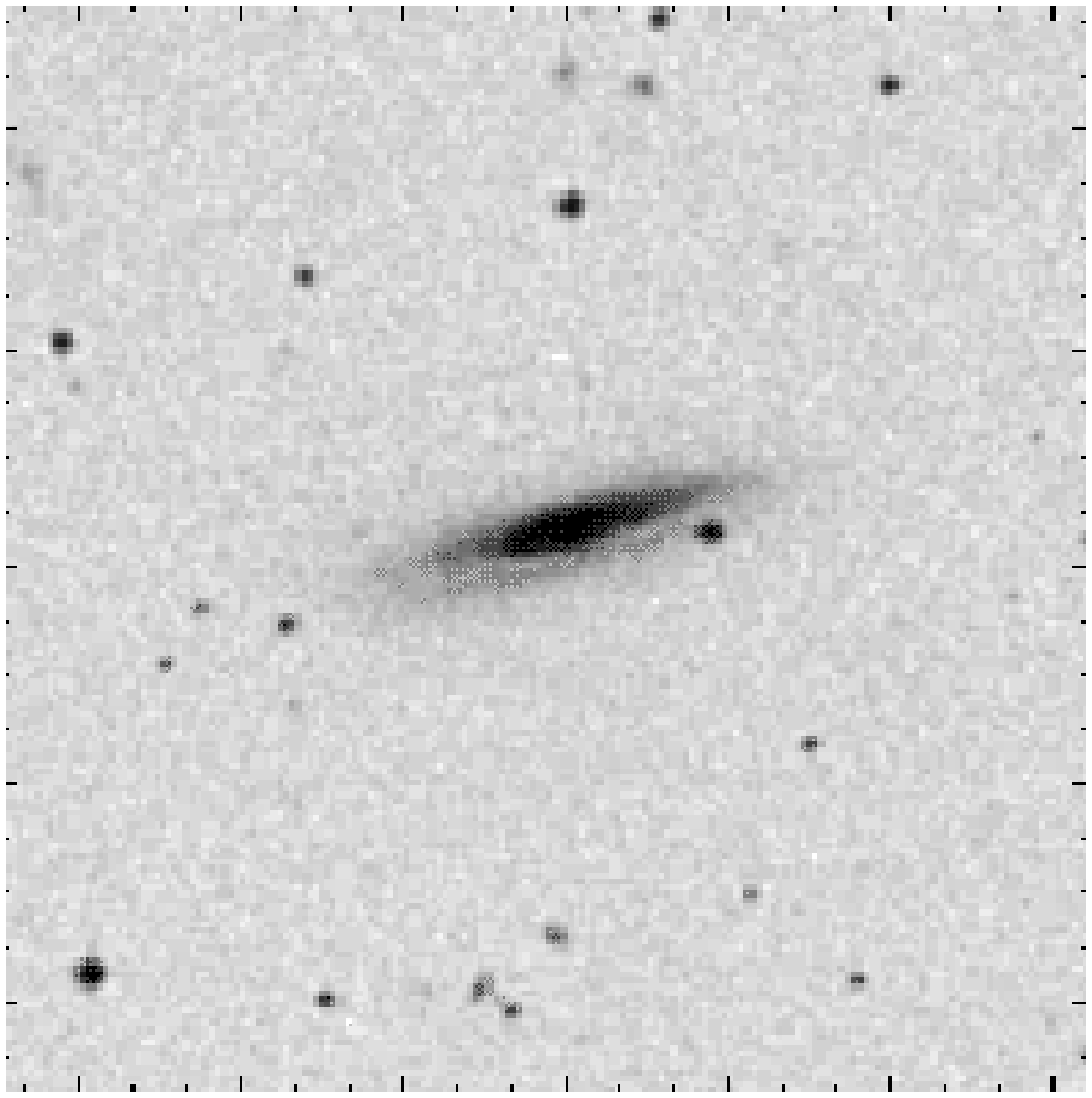"}
\special{hscale=20 vscale=20 hoffset=140 voffset=50
hsize=500 vsize=800 angle=0 psfile="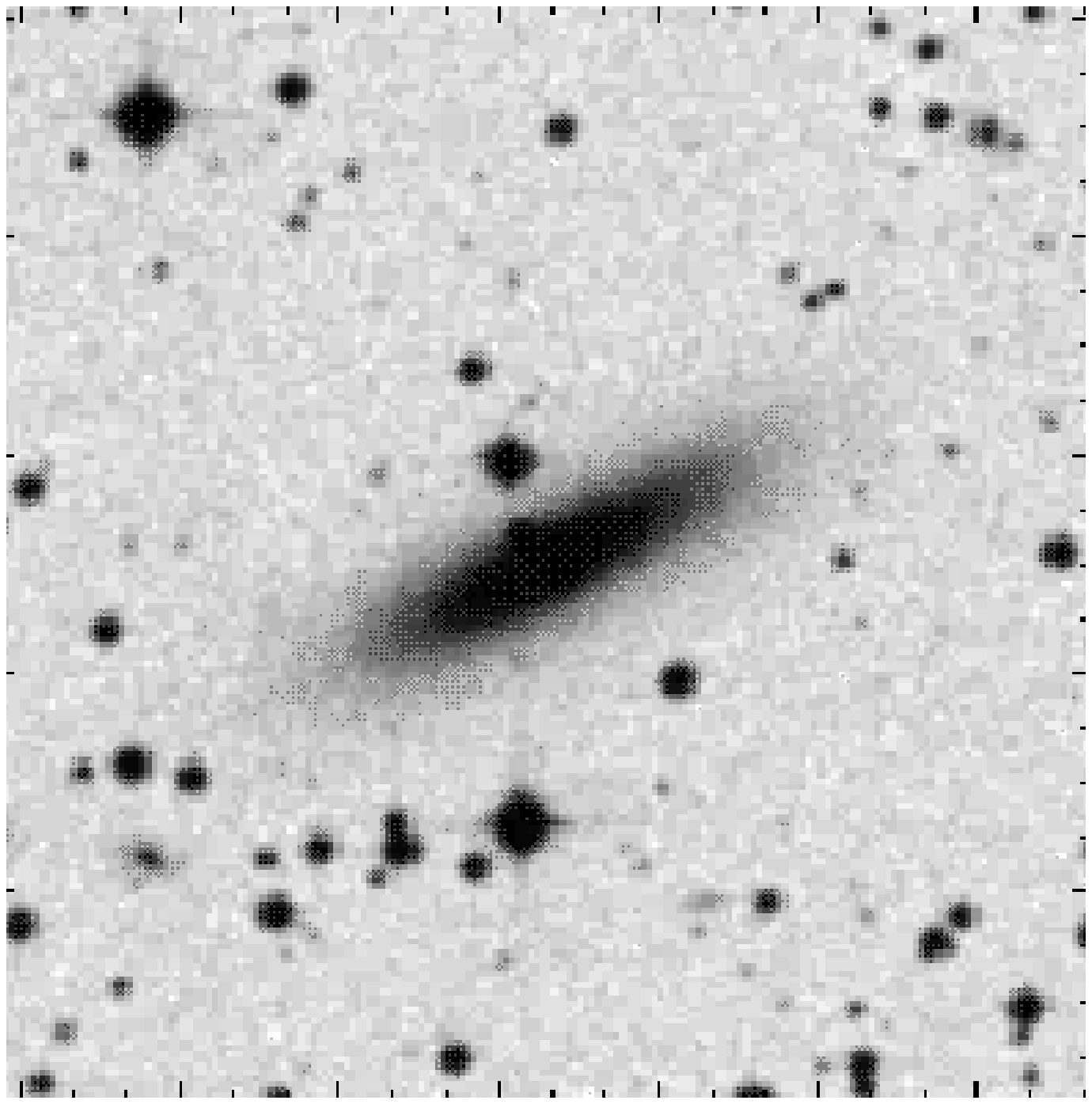"}
\special{hscale=20 vscale=20 hoffset=-20 voffset=-30
hsize=500 vsize=800 angle=0 psfile="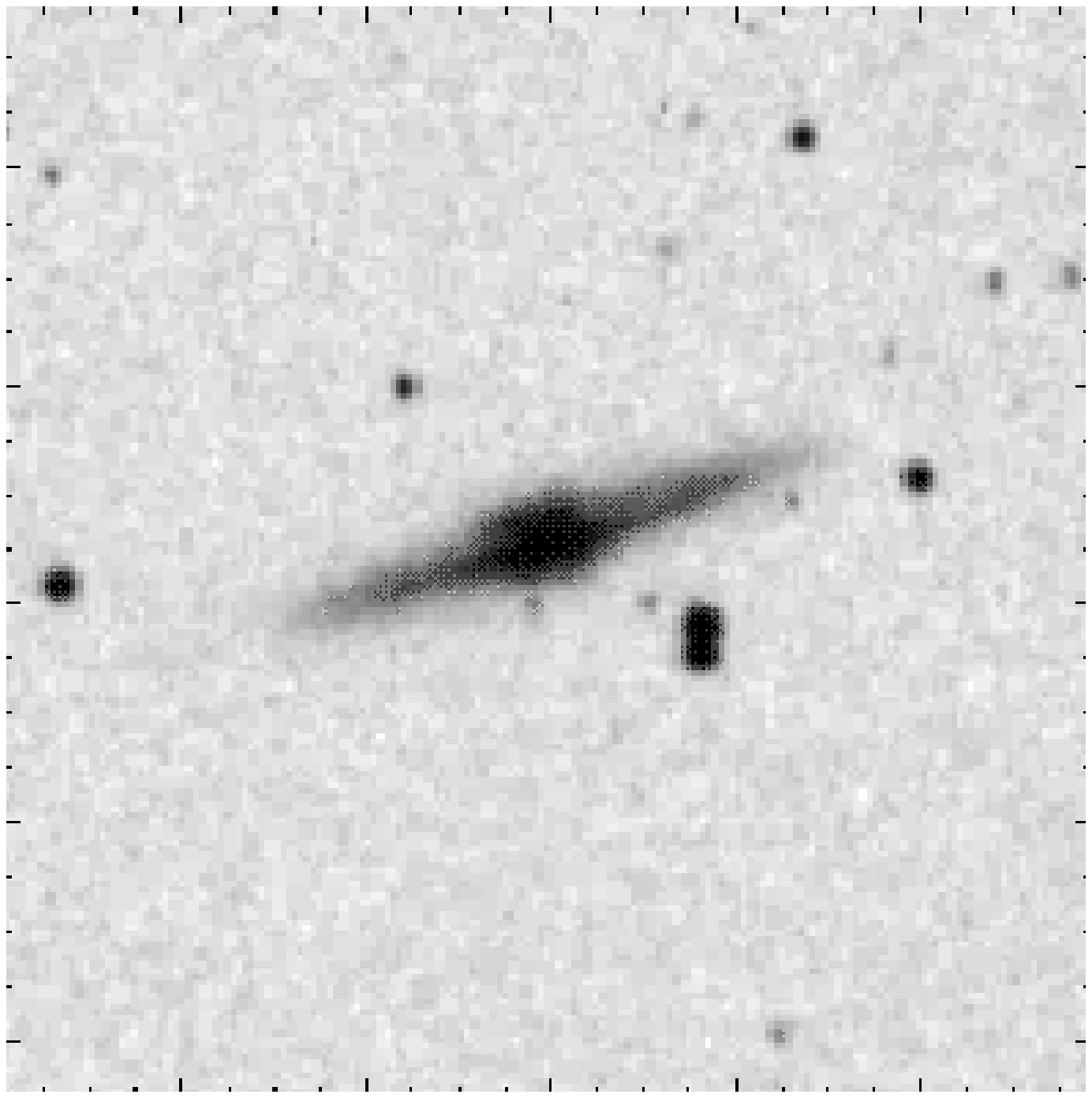"}
\special{hscale=20 vscale=20 hoffset=60 voffset=-30
hsize=500 vsize=800 angle=0 psfile="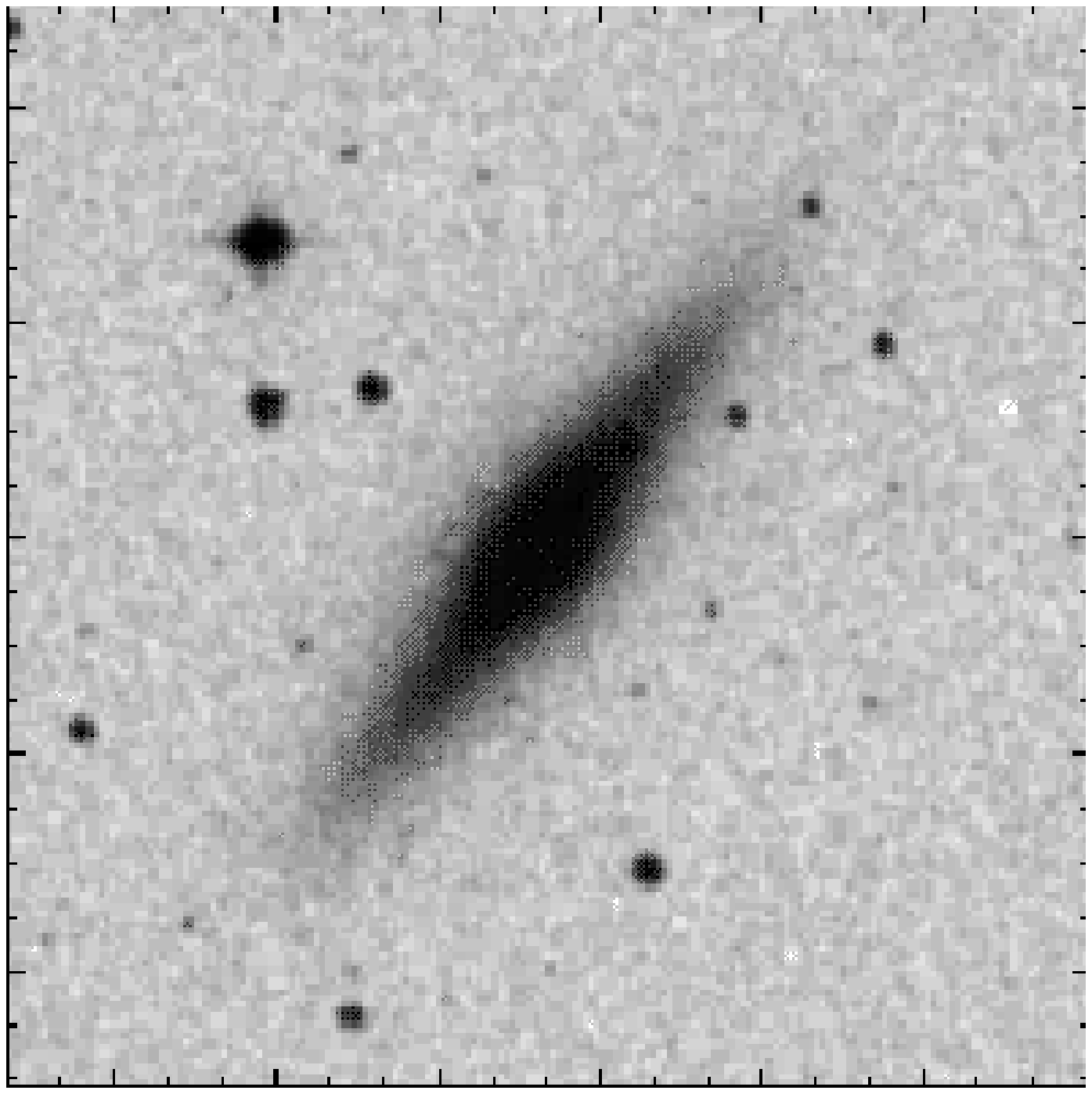"}
\special{hscale=20 vscale=20 hoffset=140 voffset=-30
hsize=500 vsize=800 angle=0 psfile="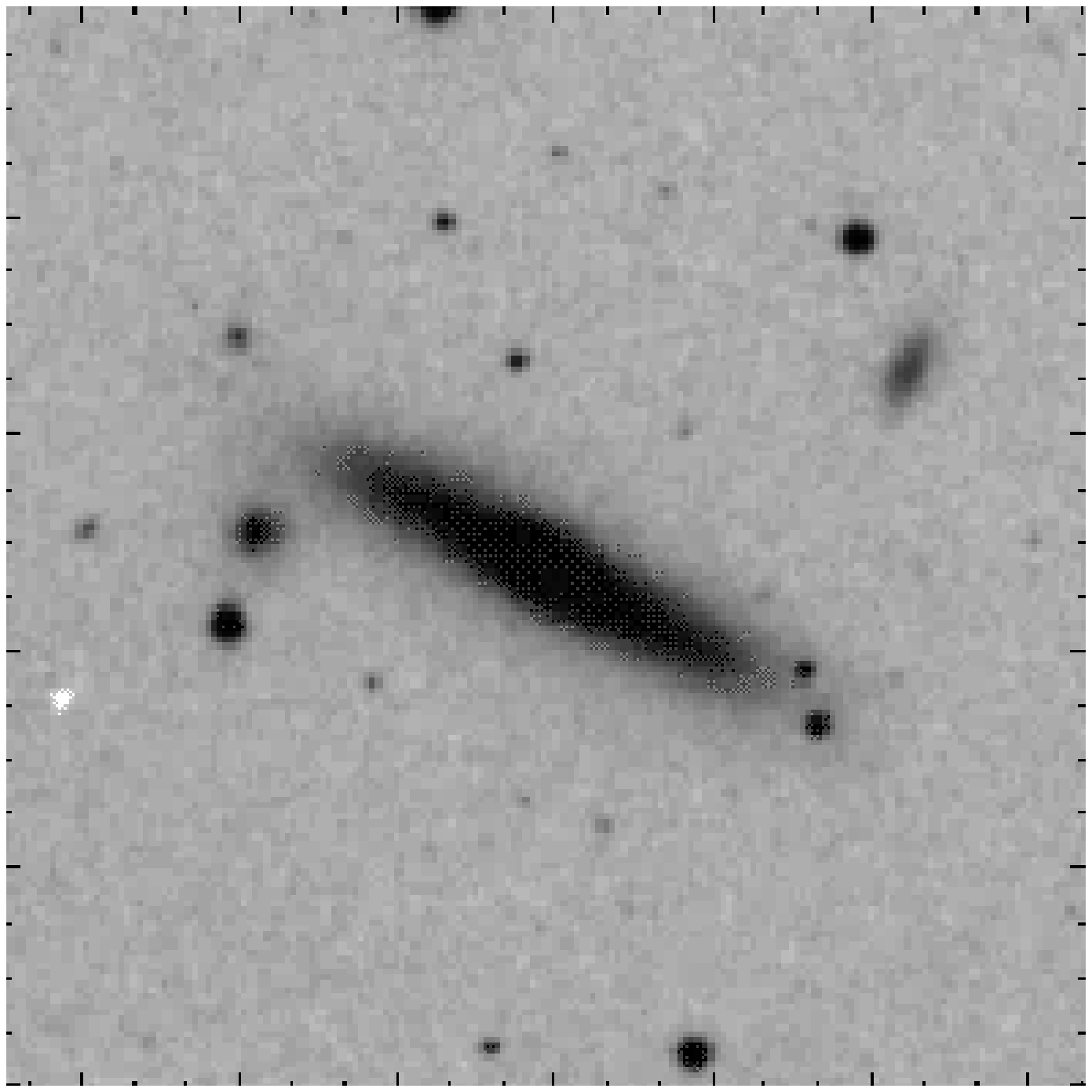"}
\caption{Images from Digitized Sky Survey for the sample of S0 
galaxies. From
Left to Right, Top to Bottom: NGC1184, NGC1611, NGC2612, NGC3896,
NGC4179 and NGC5308.}  
\label{sampleimages}
\end{figure}

The galaxies chosen for this study were selected from the sample
presented in Kuijken, Fisher \& Merrifield (1996). This galaxy sample
contains 28 early-type disc galaxies, and was originally observed to
search for counter-rotating populations, so high signal-to-noise ratio
spectra were obtained.  Details of the spectroscopic observations and
standard data reduction can be found in Kuijken, Fisher \& Merrifield
(1996).  From this sample we selected the six S0 galaxies of with
similar mean rotation speeds, so that there is some chance that the
systems are comparable in their dynamics.  We also selected galaxies
that are close enough to edge-on for it to be apparent that the
systems are disc-dominated, with relatively small bulges (and thus
that a simple evolutionary path from later-type spiral galaxies is
plausible).  Images of the sample galaxies are shown in
Fig.~\ref{sampleimages}.

\subsection{Kinematic analysis}
LOSVDs were extracted from the reduced spectra using the unresolved
gaussian decomposition (UGD) algorithm (Kuijken \& Merrifield 1993).
This method models each LOSVD as the sum of a set of unresolved
gaussians with fixed means and dispersions.  The amplitudes of the
gaussians are varied so as to minimize the difference between the
galaxy spectrum and the model derived by convolving the LOSVD with a
suitable template star spectrum. This approach has the advantage that
it does not force any particular functional form on the LOSVD, beyond
the smoothness imposed by the widths of the component gaussians.  It
is therefore well-suited to modeling complex systems like edge-on disc
galaxies, where integration along the line of sight, potentially
through multiple components, can make the shape of the LOSVD rather
complex.  

\begin{figure}
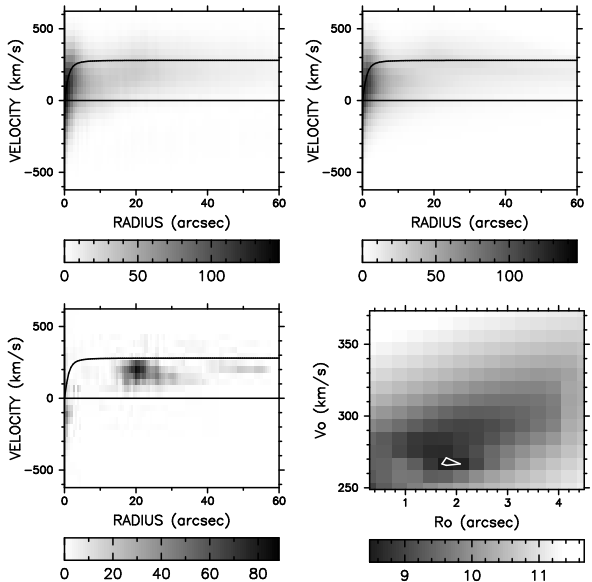

\vspace*{8.5cm}
\special{hscale=40 vscale=40 hoffset=0 voffset=240
hsize=500 vsize=800 angle=-90 psfile="ngc1184plot.ps"}
\special{hscale=40 vscale=40 hoffset=115 voffset=127 hsize=500
vsize=800 angle=-90 psfile="ngc1184chi2.ps"} 

\caption{Top left: The line-of-sight velocity distribution as a
function of projected radius along the major axis of NGC~1184.  Top
right: the corresponding quantities for the best-fit dynamical model.
Bottom left: plot of $\mbox{[(data-model)/}\sigma\mbox{]}^2$, showing
where the major $\chi^2$ residuals arise.  Bottom right: $\ln(\chi^2)$
as a function of the adopted parameters for the gravitational
potential, $(r_0,v_0)$; the solid contour shows where $\ln(\chi^2) =
\ln(1.1 \times \chi^2_{min})$.}
\label{ngc1184plot}
\end{figure}

\begin{figure}
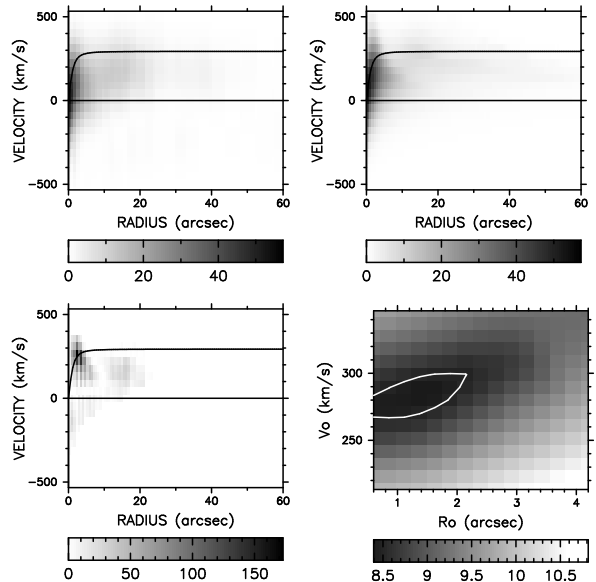

\vspace*{8.5cm}
\special{hscale=40 vscale=40 hoffset=0 voffset=240
 hsize=500 vsize=800 angle=-90 psfile="ngc1611plot.ps"}
\special{hscale=40 vscale=40 hoffset=115 voffset=127 hsize=500 
vsize=800 angle=-90 psfile="ngc1611chi2.ps"} 
\caption{Same as in Fig.~\ref{ngc1184plot} for NGC~1611}   
\label{ngc1611plot}
\end{figure}

\begin{figure}
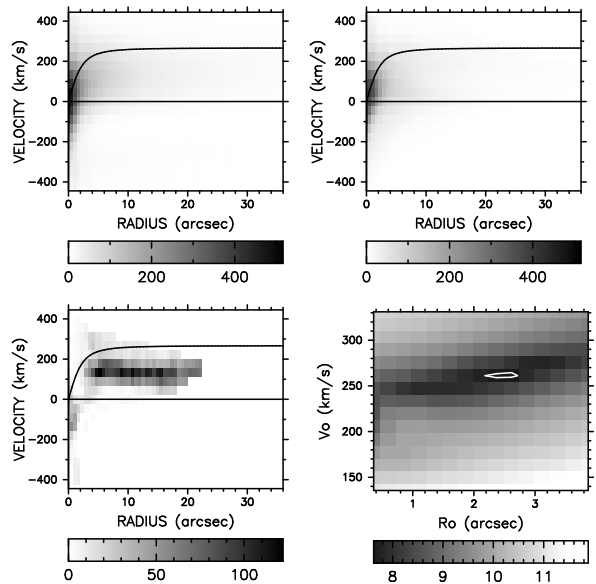

\vspace*{8.5cm}
\special{hscale=40 vscale=40 hoffset=0 voffset=240
hsize=500 vsize=800 angle=-90 psfile="ngc2612plot.ps"}
\special{hscale=40 vscale=40 hoffset=115 voffset=127 hsize=500 vsize=800 angle=-90 psfile="ngc2612chi2.ps"} 
\caption{Same as in Fig.~\ref{ngc1184plot} for NGC~2612}   
\label{ngc2612plot}
\end{figure}

\begin{figure}
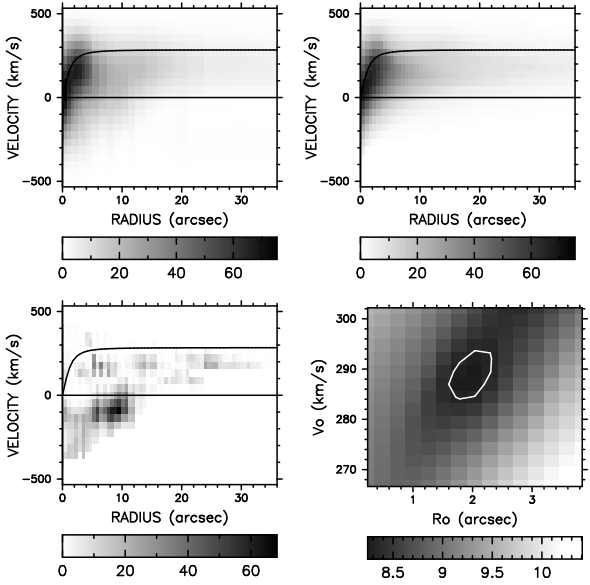

\vspace*{8.5cm}
\special{hscale=40 vscale=40 hoffset=0 voffset=240
hsize=500 vsize=800 angle=-90 psfile="ngc3986plot.ps"}
\special{hscale=40 vscale=40 hoffset=115 voffset=127 hsize=500
vsize=800 angle=-90 psfile="ngc3986chi2.ps"} 
\caption{Same as in Fig.~\ref{ngc1184plot} for NGC~3986}   
\label{ngc3986plot}
\end{figure}

\begin{figure}
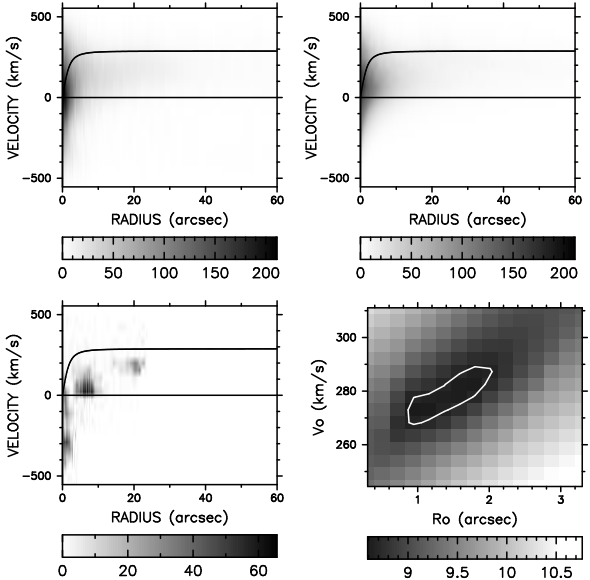

\vspace*{8.5cm}
\special{hscale=40 vscale=40 hoffset=0 voffset=240
hsize=500 vsize=800 angle=-90 psfile="ngc4179plot.ps"}
\special{hscale=40 vscale=40 hoffset=115 voffset=127 hsize=500
vsize=800 angle=-90 psfile="ngc4179chi2.ps"} 
\caption{Same as in Fig.~\ref{ngc1184plot} for NGC~4179}   
\label{ngc4179plot}
\end{figure}

\begin{figure}
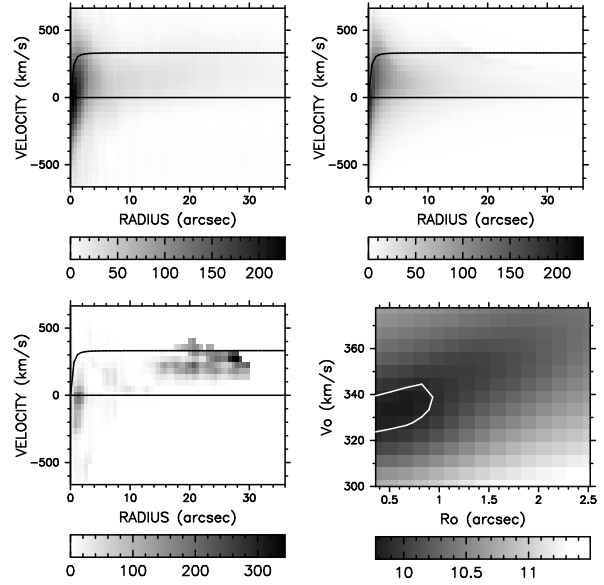

\vspace*{8.5cm}
\special{hscale=40 vscale=40 hoffset=0 voffset=240
hsize=500 vsize=800 angle=-90 psfile="ngc5308plot.ps"}
\special{hscale=40 vscale=40 hoffset=115 voffset=127 hsize=500
vsize=800 angle=-90 psfile="ngc5308chi2.ps"} 
\caption{Same as in Fig.~\ref{ngc1184plot} for NGC~5308}   
\label{ngc5308plot}
\end{figure}

In order to increase the signal-to-noise ratio of the resulting
kinematic data, the two-dimensional LOSVD as a function of position
along the major axis, $F(v_{\rm los}, R)$, was assumed to have the
symmetry of an edge-on axisymmetric disc system, so that $F(v_{\rm
los}, R) \equiv F(-v_{\rm los}, -R)$.  We therefore determined the
kinematic coordinates of the centre of each galaxy by finding the
point in $F(v_{\rm los}, R)$ at which this symmetry is most closely
obeyed.  We then averaged the data from the two sides of each galaxy.
The resulting mean estimates for $F(v_{\rm los}, R)$ are shown in the
top left panels of Figs.~\ref{ngc1184plot}, \ref{ngc1611plot},
\ref{ngc2612plot}, \ref{ngc3986plot}, \ref{ngc4179plot} and
\ref{ngc5308plot}. 

\subsection{Dynamical analysis}
Having derived these kinematic data, we must now attempt to describe
them using a dynamical model.  The simplest credible model for the
major axis kinematics of a galaxy involved fitting them to a
two-integral distribution function, $f(E,L)$, which describes the
phase-space density of stars, where $E$ is the energy and $L$ is the
angular momentum of the stars about the symmetry axis, two of the
integrals of motion (Binney \& Tremaine 1987).  Such a model would be
exact for an infinitely-thin axisymmetric disc, but it is also a
credible approximation for a system of finite thickness.  Defining the
usual polar coordinates, the energy in the $z$ direction, $E_z =
\Phi(R,z) - \Phi(R,0) + {1\over2}v_z^2$, is approximately an integral
of motion for a thin disc.  A system whose distribution function takes
the form $f(E,L)g(E_z)$ will have dynamics in the $z=0$ plane
identical to those of an infinitely-thin disc with a distribution
function $f(E,L)$.  Thus, the dynamics in the plane of the disc can
credibly be modeled by treating the major-axis kinematics as if they
were those of an infinitely-thin disc.

Mathieu \& Merrifield (2000) presented an algorithm by which the
observed density of stars as a function of projected radius and
line-of-sight velocity, $F(R, v_{\rm los})$ could be iteratively
inverted to find the distribution function, $f(E,L)$, that would
produce such observable kinematics.  This method builds an estimate of
the distribution function using just the data from the upper part of
the line-of-sight velocity distribution where the velocities exceed
the circular velocity.  If the correct gravitational potential is
adopted, then the rest of the velocity distribution automatically
matches the data; however, a mismatch will occur if the wrong
potential is assumed.  Thus, this algorithm algorithm returns not only
an estimate for the disc distribution function, but also constrains
the form of the gravitational potential.  

In principle, the gravitational potential in the disc plane could take
any form, but the data are not of high enough quality to allow a
completely non-parametric derivation.  We therefore adopt the simple
but sufficiently general form of a softened isothermal sphere
potential of the form
\be 
\Phi(r) = \frac{v^2_0}{2} \ln \left( 1 + \frac{r^2}{r^2_0} \right).
\label{Psidef}
\ee 
For this potential, the circular velocity can be written
\be
v_{c}(r) = \frac{v_0\, r}{\sqrt{r^2+r_0^2}}, 
\ee
with the parameters $r_0$ and $v_0$ specifying how quickly the
rotation curve rises and its asymptotic value.  

\begin{table}
\begin{array}[t]{lccccc}
{\rm Name}  & v_0\ [km/s] & r_0\ [arcsec] & r_d\ [arcsec] & M_I & M_H\\
\hline
\hline
NGC~1184 &  271 \pm 17 & 1.4 \pm 0.4 & 21.3 \pm 1.5 & -21.1 & -23.0\\
NGC~1611 & 271 \pm 15 & 1.8 \pm 0.3 & 12.1 \pm 0.3 & -21.5 & -23.4 \\
NGC~2612 & 252 \pm 11 & 2.2 \pm 0.3 & 12.8 \pm 0.2 & -19.9 & -21.8 \\
NGC~3986 & 289 \pm 06 & 2.4 \pm 0.6 & 07.6 \pm 1.1 & -21.6 & -23.5\\
NGC~4179 & 275 \pm 07 & 1.7 \pm 0.4 & 12.8 \pm 0.5 & -20.9 & -22.8\\
NGC~5308 & 321 \pm 17 & 0.5 \pm 0.1 & 16.5 \pm 0.4  & -21.6 & -23.5\\
\hline
\end{array}
\caption{Estimates of the parameters $(r_0,v_0)$ for the gravitational
potential, estimate of the photometric disc scale length $r_d$ and
absolute I and H magnitudes for each galaxy.}
\label{table1}
\end{table}

\begin{figure*}
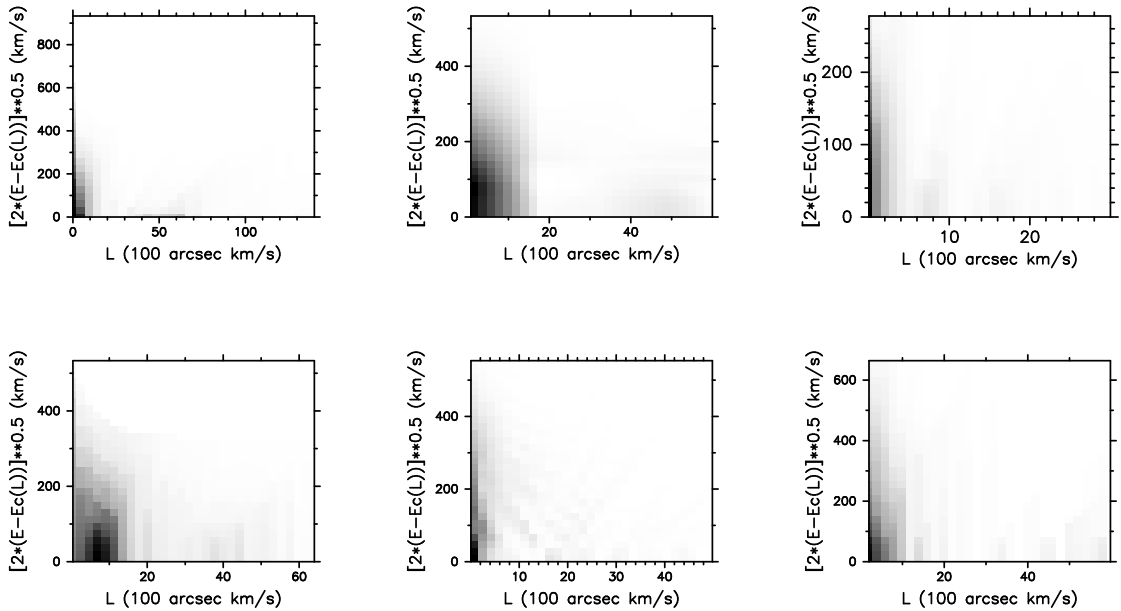

\vspace*{8.5cm}
\special{hscale=45 vscale=45 hoffset=30 voffset=250
hsize=500 vsize=800 angle=-90 psfile="ngc1184df.ps"}
\special{hscale=45 vscale=45 hoffset=180 voffset=250
hsize=500 vsize=800 angle=-90 psfile="ngc1611df.ps"}
\special{hscale=45 vscale=45 hoffset=330 voffset=250
hsize=500 vsize=800 angle=-90 psfile="ngc2612df.ps"}
\special{hscale=45 vscale=45 hoffset=30 voffset=120
hsize=500 vsize=800 angle=-90 psfile="ngc3986df.ps"}
\special{hscale=45 vscale=45 hoffset=180 voffset=120
hsize=500 vsize=800 angle=-90 psfile="ngc4179df.ps"}
\special{hscale=45 vscale=45 hoffset=330 voffset=120
hsize=500 vsize=800 angle=-90 psfile="ngc5308df.ps"}
\caption{The disc distribution functions for the galaxies as derived
iteratively from their line-of-sight kinematics using the best-fit
gravitational potentials.}
\label{dffig}
\end{figure*}

Using the iterative algorithm, distribution functions have been
derived for all the sample galaxies.  Figure~\ref{dffig} shows the
best-fit distribution functions obtained in this way.  All the systems
show the single concentration of low angular-momentum material,
characteristic of a disc.  Thus, there is no evidence even in these
detailed dynamical models of any peculiarities in the stellar dynamics
that one might expect if these systems had formed in a spectacular
manner such as in a merger (Bekki 1998).

To assess the reliability of the fitting process,
Figs.~\ref{ngc1184plot}, \ref{ngc1611plot}, \ref{ngc2612plot},
\ref{ngc3986plot}, \ref{ngc4179plot} and \ref{ngc5308plot} show the
``observable'' kinematics of these dynamical model, $F(R, v_{\rm
los})$, and the difference between these predictions and the observed
kinematics (normalized by the uncertainty in the observations).
Finally, these figures also show how rapidly the goodness of fit (in a
$\chi^2$ sense) degrades as one goes away from the optimal values for
$r_0$ and $v_0$ in the gravitational potential.  The best-fit values
for $r_0$ and $v_0$ are listed in Table~\ref{table1}.

\begin{figure}
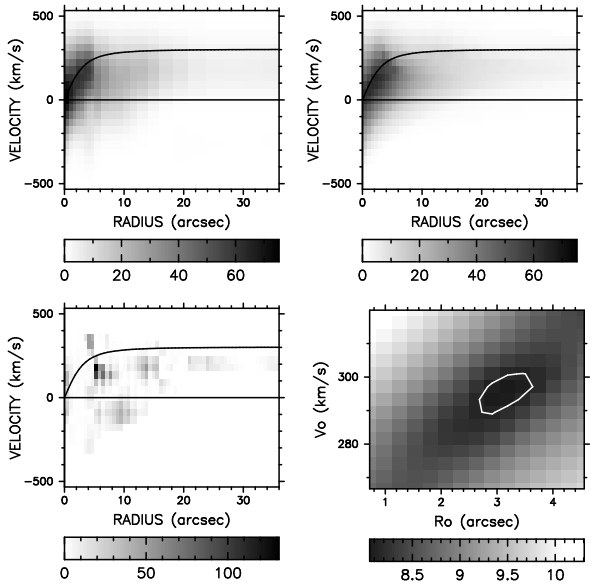

\vspace*{8.5cm}
\special{hscale=40 vscale=40 hoffset=0 voffset=240
hsize=500 vsize=800 angle=-90 psfile="ngc3986leftplot.ps"}
\special{hscale=40 vscale=40 hoffset=115 voffset=127 hsize=500
vsize=800 angle=-90 psfile="ngc3986leftchi2.ps"} 
\caption{Same as in Fig.~\ref{ngc1184plot} for NGC~3986, using data on
the "left" side of the galaxy.}
\label{ngc3986left}
\end{figure}

\begin{figure}
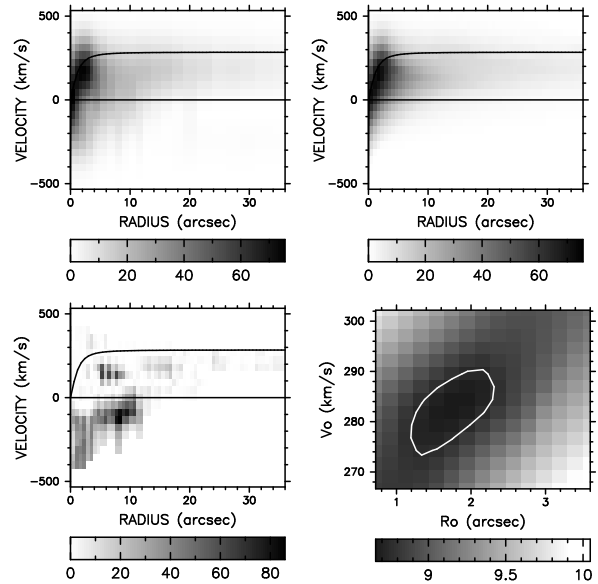

\vspace*{8.5cm}
\special{hscale=40 vscale=40 hoffset=0 voffset=240
hsize=500 vsize=800 angle=-90 psfile="ngc3986rightplot.ps"}
\special{hscale=40 vscale=40 hoffset=115 voffset=127 hsize=500
vsize=800 angle=-90 psfile="ngc3986rightchi2.ps"} 
\caption{Same as in Fig.~\ref{ngc1184plot} for NGC~3986, using data on
the "right" side of the galaxy.}
\label{ngc3986right}
\end{figure}

So far in our analysis we assumed that the galaxies are perfectly
axisymmetric and we used average data from both sides of the
galaxy. However the kinematics of these galaxies show small
asymmetries and we also performed the fitting process on each side of
the galaxies separately. We use the best-fit values of $r_0$ and $v_0$
on each side of the galaxy to estimate the dispersion of $r_0$ and
$v_0$ around the mean for each galaxy. An example of fits on both
sides of one galaxy (NGC3986) is shown in Figs.~\ref{ngc3986left} and
~\ref{ngc3986right}.

\begin{figure}
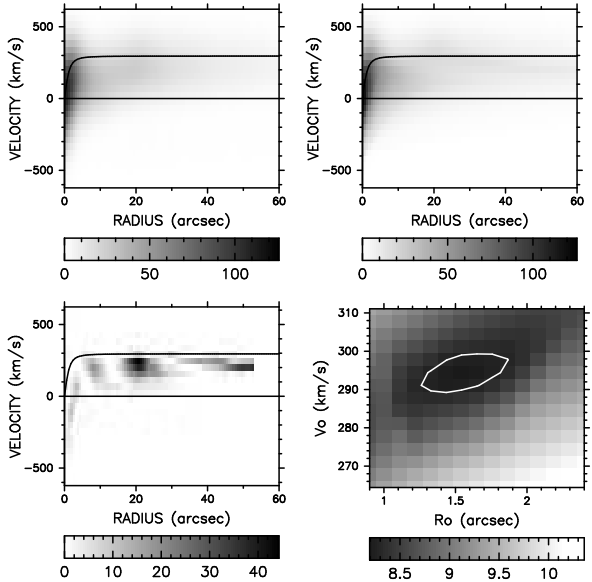

\vspace*{8.5cm}
\special{hscale=40 vscale=40 hoffset=0 voffset=240
hsize=500 vsize=800 angle=-90 psfile="ngc1184bplot.ps"}
\special{hscale=40 vscale=40 hoffset=115 voffset=127 hsize=500
vsize=800 angle=-90 psfile="ngc1184bchi2.ps"} 
\caption{Same as in Fig.~\ref{ngc1184plot} for a bulge-disc model of a
NGC~1184 }
\label{ngc1184bplot}
\end{figure}

Clearly, in these galaxies there are some systematic residuals in the
difference between model and data.  Indeed, the minimum values of
$\chi^2$ are well above what one might expect for a formally good fit.
However, given the assumptions involved as to the separable form of
the distribution function and the adopted parametric function for the
gravitational potential, these small residuals are not particularly
surprising.  In order to assess their impact on our subsequent
analysis, we have explored more sophisticated models for the
distribution function.  In particular, we have investigated what
happens if our assumption of a single-component disc-like distribution
function is relaxed, by explicitly adding the contribution from a
central bulge component (rather than treating it as part of the disc).
To this end, we adopt a simple bulge model with a distribution
function of the form
\begin{equation}
f(E,L) = \alpha [ \exp (-\beta E ) - \exp( -\beta E_0) ] \exp (\gamma L), 
\end{equation}
where $(E \leq E_0)$ is the energy of a star and $L$ its total angular
momentum, and where $\alpha$, $\beta$ and $\gamma$ are constants (see
Jarvis \& Freeman, 1985).  We then computed the line-of-sight velocity
distribution corresponding to this distribution function and subtract
it from the observed line-of-sight velocity distribution.  The choice
of the constants remains somewhat arbitrary as a full bulge-disc
decomposition is underconstrained, but we stick to values of the
parameters such that the estimate of line-of-sight velocity
distribution of the disc alone (after subtracting the line-of-sight
velocity distribution produced by the bulge from the observed
line-of-sight velocity distribution) is positive almost everywhere.
Rather surprisingly, this extra spherical component makes very little
difference to the best-fit gravitational potential:
Fig.~\ref{ngc1184bplot} illustrates the best-fit model for NGC~1184,
in which the constraints on $r_0$ and $v_0$ are close 
to those in the pure disc model.  It would thus appear that the simple
model adopted above does a robust job of estimating the gravitational
potential even where the assumption of a thin disc is not strictly
valid.  

\subsection{Photometric analysis}

B-band integrated magnitudes for the sample galaxies have been taken
from the RC3 catalogue, where available, or from the Lyon-Meudon
extragalactic database (LEDA) otherwise.  These have been converted to
absolute magnitudes using distance estimates from the LEDA database,
where available. For NGC1611 and NGC2612, as no distance is available,
we estimate the distance using the systemic velocity derived from our
spectra assuming $H_0=80\ km s^{-1} Mpc^{-1}$. It would be useful to
study the integrated photometric properties of these systems in a
number of bands.  Unfortunately, such data are unavailable for most
systems.  However, the spread in the integrated colors of S0 galaxies
is very small: for the S0 galaxies in the sample of Persson, Frogel \&
Aaronson (1979), $B-H$ has a mean value of 3.9 and a scatter of only
$\sim 0.3$, while $B-I$ has a mean value of 2 and a scatter of $\sim
0.3$ for the S0 galaxies in the sample of Neistein et al. (1999).  As
we shall see below, these scatters are comparable to other sources of
uncertainty, so for this analysis we can simply apply the mean color
corrections in order to estimate absolute magnitudes in the $I$ and
$H$ bands. The resulting magnitudes are listed in Table~\ref{table1}.

\begin{figure*}
\vspace*{9cm}
\special{hscale=60 vscale=60 hoffset=70 voffset=-200
hsize=500 vsize=800 angle=0 psfile="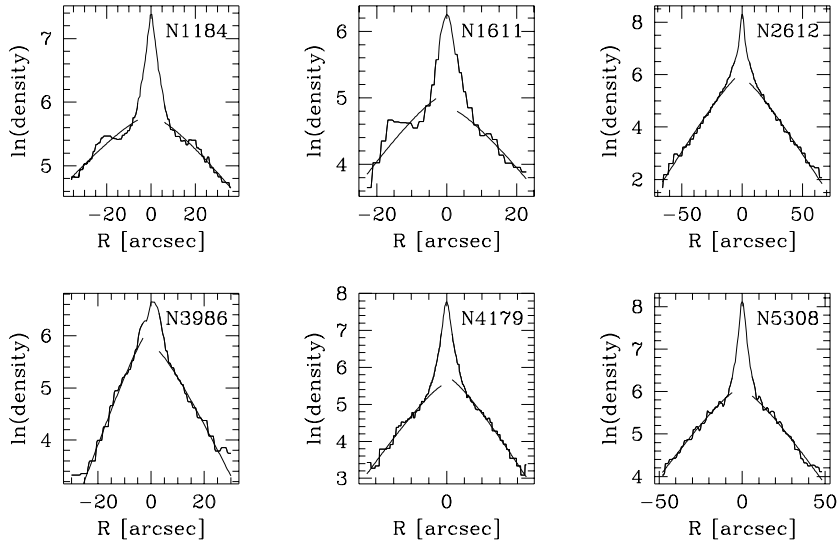"}
\caption{Plot of the logarithm of the projected luminosity density as
a function of projected radius along the major axes of the six sample
galaxies, as derived from their line-of-sight velocity distributions.
The overplotted smooth lines show the best-fit disc model.}
\label{plotphotom}
\end{figure*}

The principal spatially-resolved photometric quantity of interest in
these disc-dominated systems is the disc scale-length.  To determine
this quantity, we first estimate the intensity profiles, $I(r)$, as a
function of radius by integrating the line-of-sight velocity
distributions derived above with respect to velocity.  We adopt this
approach in preference to using imaging data, as it guarantees that
the photometric and kinematic measurements are probing the same
stellar population.  We then approximate the starlight of the disc at
large radius by fitting a function of the form

\begin{equation}
I_{\rm mod}(r) = 2 A_d |r| K_1(|r|/r_d), 
\label{densityphotom}
\end{equation}

where $A_d$ and $r_d$ are free parameters, and $K_1$ is a modified
Bessel function, which represents the projected density profile of an
edge-on exponential disc.  With only rather noisy photometric profiles
and no detailed surface photometry for our galaxies, we only fit a
disc model at large radius as fitting a bulge-disc model proves very
much underconstrained and does not give reliable values for the
parameters of the disc and bulge.  The best fit photometric models are
shown in Fig.~\ref{plotphotom} where the models are fitted
independently on each side of the galaxy, and the derived photometric
scale-lengths $r_d$ are given in Table~\ref{table1}. This quantity
differs from the photometric scale length usually estimated by fitting
two-component bulge-disc models .

\section{Discussion}\label{discussionsec}

\begin{figure}
\vspace*{6cm}
\special{hscale=30 vscale=30 hoffset=30 voffset=-50
hsize=500 vsize=800 angle=0 psfile="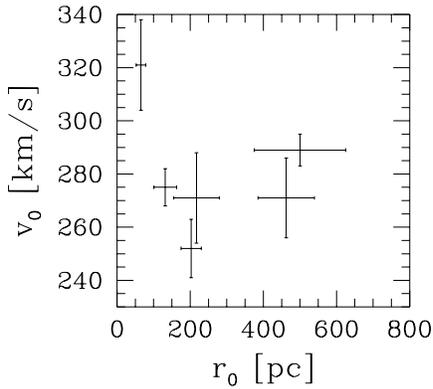"}
\caption{Plot of the estimates of the maximum circular velocity $v_0$
vs. the estimates of the parameter $r_0$.}
\label{r0v0}
\end{figure}

\begin{figure}
\vspace*{6cm}
\special{hscale=30 vscale=30 hoffset=30 voffset=-50
hsize=500 vsize=800 angle=0 psfile="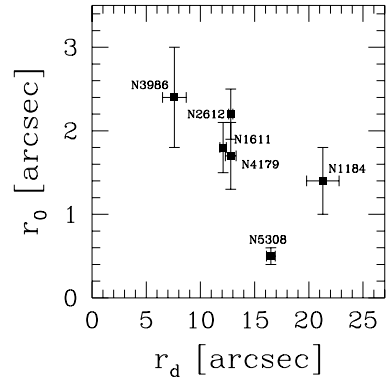"}
\caption{Plot of the estimates of photometric disc scale length $r_d$
vs. the estimates of the parameter $r_0$ for the gravitational
potential.}
\label{r0rd}
\end{figure}

Having derived the dynamical and photometric properties of these
galaxies, we are now in a position to look for clues to their origins.
Figure~\ref{r0v0} shows the physical parameters of the rotation curves
in these galaxies.  There are no signs of any correlation between
$r_0$ and $v_0$, suggesting that these galaxies occupy dark halos with
the usual wide range of characteristics, although the sample is really
too small to say anything definitive.  Similarly, there is no
discernable correlation between the mass scale-length, $r_0$, and the
photometric scale-length, $r_d$, (see Fig.~\ref{r0rd}) arguing against
Neistein {\it et al.}'s (1999) suggestion that these galaxies should
be more dominated by the mass in their stellar discs than normal
spiral galaxies.

\begin{figure}
\vspace*{6cm}
\special{hscale=30 vscale=30 hoffset=30 voffset=-50
hsize=500 vsize=800 angle=0 psfile="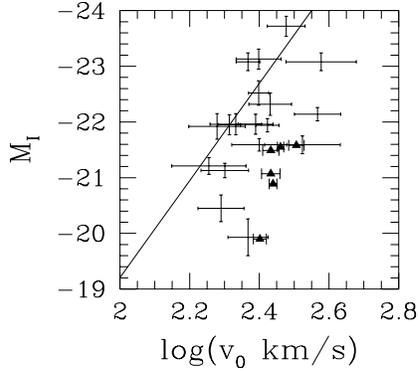"}
\caption{Absolute $I$-band magnitude versus $\log(v_0/{\rm km/s})$ for
the sample of S0 galaxies in Neistein {\it et al.}\ (1999), and those
in this analysis (filled triangles).  The line shows the $I$-band
Tully-Fisher relation for later-type spirals from Pierce \& Tully
(1992).}
\label{IbandTFfig}
\end{figure}

A more interesting result comes when we look at the Tully-Fisher
relation for S0 galaxies.  Figure~\ref{IbandTFfig} shows the I-band
relation derived by Neistein {\it et al.}\ (1999), which lie quite
close to the Tully-Fisher relation for normal spiral galaxies (Pierce
\& Tully 1992), but with a large RMS scatter of $\sim 0.6$ magnitudes
about the best-fit line parallel to the standard Tully-Fisher
relation.  The galaxies in the current sample lie systematically below
the Neistein {\it et al.}  galaxies.  With the galaxies all selected
to have comparable rotation speeds, we clearly cannot derive a slope
for the Tully-Fisher relation from these data, but fixing the slope to
that of the relation for normal spiral galaxies, we find that the
best-fit line is offset from the spiral galaxy relation by $\sim 1.8$
magnitude, with a scatter about the best-fit line of only $\sim 0.3$
magnitudes. For this fit we exclude NGC2612 as its distance estimate
is rather uncertain due to a small systemic velocity. Even allowing
for the small sample size, this scatter is smaller than that in the
Neistein {\it et al.} sample at a statistically significant level.

This small scatter and offset in the Tully-Fisher relation by $\sim
1.8$ magnitude in the $I$-band is what one would expect if star
formation had been suddenly switched off a few Gyrs ago so that these S0
galaxies contained just the old stellar population of a normal spiral
galaxy. Using stellar population models (Charlot \& Bruzual 1991),
assuming S0s galaxies had had similar formation histories as late-type
spirals until a few Gyrs ago, we would expect the stellar population of
S0s to have faded significantly, resulting in a noticeable offset in I
magnitude, such as the one observed in our sample.

\begin{figure}
\vspace*{6cm}
\special{hscale=30 vscale=30 hoffset=30 voffset=-50
hsize=500 vsize=800 angle=0 psfile="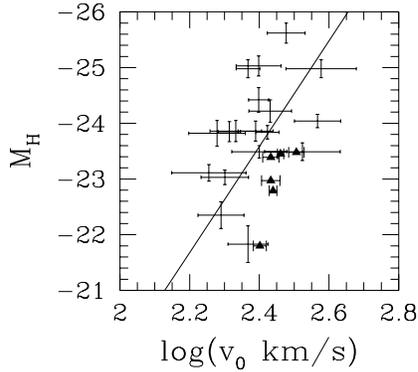"}
\caption{Estimated absolute $H$-band magnitude versus $\log(v_0/{\rm
km/s})$ for the sample of S0 galaxies in Neistein {\it et al.}\
(1999), and those in this analysis (filled triangles).  The line shows
the $H$-band Tully-Fisher relation for later-type spirals from Pierce
\& Tully (1992).}
\label{HbandTFfig}
\end{figure}

This point is made even more dramatically if we convert to estimated
$H$-band magnitudes using the prescription in Section~\ref{datasec}.
In this band, the luminosity is dominated by the old stellar
populations; as Fig.~\ref{HbandTFfig} shows, the S0s in the current
sample lie quite close to the Tully-Fisher relation for later-type
spiral galaxies, suggesting that their old stellar populations are
rather similar.

\section{Conclusions}\label{conclusionsec}
In this paper, we have calculated the first detailed dynamical models
for a small sample of edge-on S0 galaxies with small bulges.  In
addition to producing distribution functions for these disc-dominated
systems, which look very much as one would expect for normal disc
systems, the analysis also returned estimates for the parameters of
their gravitational potentials.

The interpretation of these data all points to a simple picture in
which these systems were formed by the stripping of gas from normal
spiral galaxies.  The distribution functions are all well modeled by
unexceptional stellar discs, similar to those expected in the old
stellar populations of spiral galaxies.  In addition, the galaxies
obey a reasonably tight Tully-Fisher relation, which is offset from
the relation for normal spiral galaxies by the amount that one would
expect if star formation had been shut off a few Gyrs
ago, so that all that remains in these systems are the rather fainter
old stellar populations.

This result appears to conflict with Neistein {\it et al.}'s (1999)
analysis, which showed a much greater scatter in the Tully-Fisher
relation with less systematic offset.  Part of the difference may be
due to the lower signal-to-noise ratio of the data in their larger
sample, which limited their ability to carry out detailed dynamical
modeling, particularly for galaxies that lie very close to edge-on.
However, there is also a systematic difference in the way that the
samples were selected: the edge-on galaxies in the analysis of this
paper were specifically chosen to contain small bulges.  This
selection criterion means that these galaxies are prime candidates to
have formed from gas-stripped spiral galaxies.  If, as Neistein {\it
et al.} suggest, S0s are a ``mixed bag'' that formed in a variety of
ways, it should come as no surprise that this particular subsample
obey a tight Tully-Fisher relation that is not seen in the general
population of S0s.  To test such hypotheses, we ultimately need data
of the quality presented in this paper for a much larger sample of
galaxies, extending both the range of absolute magnitudes and of other
parameters such as the bulge size.

\label{lastpage}
\end{document}